# Universal spectrum for DNA base C+G frequency distribution in Human chromosomes 1 to 24

## A. M. Selvam

Deputy Director (Retired)

Indian Institute of Tropical Meteorology, Pune 411 008, India

Email: amselvam@gmail.com

Web sites: http://www.geocities.com/amselvam

http://amselvam.tripod.com/index.html

## Abstract

Power spectra of human DNA base C+G frequency distribution in all available contiguous sections exhibit the universal inverse power law form of the statistical normal distribution for the 24 chromosomes. Inverse power law form for power spectra of space-time fluctuations is generic to dynamical systems in nature and indicate long-range space-time correlations. A recently developed general systems theory predicts the observed non-local connections as intrinsic to quantumlike chaos governing space-time fluctuations of dynamical systems. The model predicts the following. (1) The quasiperiodic Penrose tiling pattern for the nested coiled structure of the DNA molecule in the chromosome resulting in maximum packing efficiency. (2) The DNA molecule functions as a unified whole fuzzy logic network with ordered two-way signal transmission between the coding and non-coding regions. Recent studies indicate influence of non-coding regions on functions of coding regions in the DNA molecule.

## 1. Introduction

Spatially extended dynamical systems in nature exhibit fractal space-time fluctuations associated with inverse power law spectrum or 1/f noise, signifying long-range space-time correlations or memory, identified as self-organized criticality (Bak, 1988; Goldberger *et al.*, 2002; West, 2004; Milotti, 2004;. Wooley and Herbert, 2005; Wang et al., 2006; Li, 2007). Selfsimilar space-time fluctuations, now identified as fractals, have earlier been reported and studied in separate branches of science till very recently (1980s) when they were recognized as universal phenomena under the new science of 'nonlinear dynamics and chaos'. The multidisciplinary nature of investigations will help gain new insights and develop mathematical and statistical techniques and analytical tools for understanding and quantifying the physics of the observed long-range correlations in dynamical systems in nature. The physics of dynamical systems therefore comes under the broad category of general systems theory. The subunits of the system function as a unified whole two-way communication and control network with global (system level) control/response to local functions/stimuli, thereby possessing the criteria for a robust system (Csete and Doyle, 2002; Kitano, 2002; 2004). Kitano (2002) makes the point that robustness is a property of an entire system; it may be that no individual component or process within a system would be robust, but the system-wide architecture still provides robust behavior. This presents a challenge for analysis, since elucidating such behaviors can be counterintuitive and computationally demanding.

DNA sequences represent a condensed archive of information on the structure and function of DNA, both as a complex machinery inside the cell and as the genetic memory for the entire organism. DNA may be taken as representative of the simultaneous needs of order and plasticity of living systems. Therefore, the characterisation of the relation between DNA structure and function and the statistical properties of the distribution of its nucleotides may offer us a reliable basis for the further development of a holistic approach. The statistical properties of DNA sequences have been studied extensively in the last 15 years. The general result that emerges from these studies is that DNA statistics is characterised by short-range and long-range correlations which are linked to the functional role of the sequences. Specifically, while coding sequences seem to be almost uncorrelated, non-coding sequences show long-range power-law correlations typical of scale invariant systems (Buiatti and Buiatti, 2004).

During the late 1960s papers began appearing that showed eukaryotic DNA contained large quantities of repetitive DNA which did not appear to code for proteins. By the early 1970s, the term "junk DNA" had been coined to refer to this non-coding DNA. Junk DNA seemed like an appropriate term for DNA cluttering up the genome while contributing in no way to the protein coding function of DNA; yet there seemed to be so much of this non-coding DNA that its significance could not be ignored. Non-coding DNA makes up a significant portion of the total genomic DNA in many eukaryotes. For example, older sources estimate 97% of the human genome to be non-coding DNA, while the recently published sequence data increases the estimates to 98.9% non-coding DNA. Introns, the DNA sequences that interrupt coding sequences and do not code for proteins themselves along with other non-coding DNA, play an important role in repression of genes and the sequential switching of genes during development, suggesting that up to 15 % of "junk DNA" functions in this vital role Standish (2002).

In this paper it is shown that the spectra of human chromosomes 1 to 24 base C+G frequency distributions follows the universal inverse power law form of the statistical normal distribution consistent with predictions of a recently developed general systems theory model for dynamical systems of all space-time scales.

## 2. General systems theory concepts

In summary (Selvam, 1990; Selvam and Fadnavis, 1998), the model is based on Townsend's concept (Townsend, 1956) that large eddy structures form in turbulent flows as envelopes of enclosed turbulent eddies. Such a simple concept that space-time averaging of small-scale structures gives rise to large-scale space-time fluctuations leads to the following important model predictions.

## 2.1 Quantumlike chaos in turbulent fluid flows

Since the large eddy is but the integrated mean of enclosed turbulent eddies, the eddy energy (kinetic) distribution follows statistical normal distribution according to the Central Limit Theorem (Ruhla, 1992). Such a result, that the additive amplitudes of the eddies, when squared, represent probability distributions is found in the subatomic dynamics of quantum systems such as the electron or photon. Atmospheric flows, or, in general turbulent fluid flows follow quantumlike chaos.

## 2.2 Dynamic memory (information) circulation network

The root mean square (r.m.s.) circulation speeds $W$ and $w_*$ of large and turbulent eddies of respective radii $R$ and $r$ are related as

$$W^2 = \frac{2}{\pi}\frac{r}{R}w_*^2 \tag{1}$$

Eq.(1) is a statement of the law of conservation of energy for eddy growth in fluid flows and implies a two-way ordered energy flow between the larger and smaller scales. Microscopic scale perturbations are carried permanently as internal circulations of progressively larger eddies. Fluid flows therefore act as dynamic memory circulation networks with intrinsic long-term memory of short-term fluctuations.

## 2.3 Quasicrystalline structure

The flow structure consists of an overall logarithmic spiral trajectory with Fibonacci winding number and quasiperiodic Penrose tiling pattern for internal structure (Fig.1). Primary perturbation $OR_O$ (Fig.1) of time period $T$ generates return circulation $OR_1R_O$ which, in turn, generates successively larger circulations $OR_1R_2$, $OR_2R_3$, $OR_3R_4$, $OR_4R_5$, etc., such that the successive radii form the Fibonacci mathematical number series, i.e., $OR_1/OR_O = OR_2/OR_1 = \ldots\ldots= \tau$ where $\tau$ is the golden mean equal to $(1+\sqrt{5})/2 \approx 1.618$. The flow structure therefore consists of a nested continuum of vortices, i.e., vortices within vortices.

The quasiperiodic Penrose tiling pattern with five-fold symmetry has been identified as quasicrystalline structure in condensed matter physics (Janssen, 1988). The self-organized large eddy growth dynamics, therefore, spontaneously generates an internal structure with the five-fold symmetry of the dodecahedron, which is referred to as the icosahedral symmetry. Recently the carbon macromolecule $C_{60}$, formed by condensation from a carbon vapour jet, was found to exhibit the icosahedral symmetry of the closed soccer ball and has been named Buckminsterfullerene or footballene (Curl and Smalley, 1991). Self-organized quasicrystalline pattern formation therefore exists at the molecular level also and may result in condensation of specific biochemical structures in biological media. Logarithmic spiral formation with Fibonacci winding number and five-fold symmetry possess maximum packing efficiency for component parts and are manifested strikingly in plant Phyllotaxis (Jean, 1994).

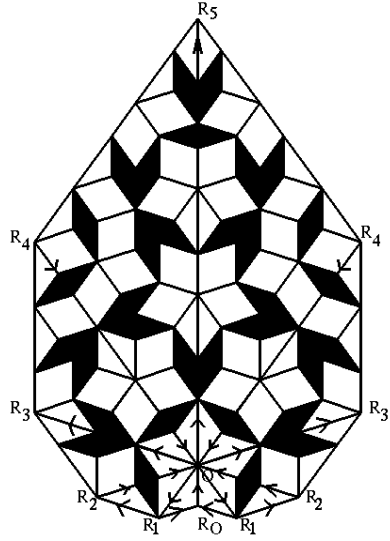

Figure 1: Internal structure of large eddy circulations. Large eddies trace an overall logarithmic spiral trajectory $OR_OR_1R_2R_3R_4R_5$ simultaneously in clockwise and anti-clockwise directions with the quasi-periodic Penrose tiling pattern for the internal structure

## 2.4 Dominant periodicities

Dominant quasi-periodicities $P_n$ corresponding to the internal circulations (Fig.1) $OR_OR_1$, $OR_1R_2$, $OR_2R_3$, ..... are given as

$$P_n = T(2 + \tau)\tau^n \qquad (2)$$

The dominant quasi-periodicities are equal to $2.2T$, $3.6T$, $5.8T$, $9.5T$, ......for values of $n = -1, 0, 1, 2,...,$ respectively (Eq.2). Space-time integration of turbulent fluctuations results in robust broadband dominant periodicities which are functions of the primary perturbation time period $T$ alone and are independent of exact details (chemical, electrical, physical etc.) of turbulent fluctuations. Persistent periodic energy pumping at fixed time intervals (period) $T$ in a fluid medium generates self-sustaining continuum of eddies and results in apparent nonlinear chaotic fluctuations in the fluid medium. Also, such global scale oscillations in the unified network are not affected appreciably by failure of localized microscale circulation networks.

Wavelengths (or periodicities) close to the model predicted values have been reported in weather and climate variability (Selvam and Fadnavis, 1998), prime number distribution (Selvam, 2001a), Riemann zeta zeros (non-trivial) distribution (Selvam, 2001b), Drosophila DNA base sequence (Selvam, 2002), stock market economics (Selvam, 2003), Human chromosome 1 DNA base sequence (Selvam, 2004).

Macroscale coherent structures emerge by space-time integration of microscopic domain fluctuations in fluid flows. Such a concept of the autonomous growth of atmospheric eddy continuum with ordered energy flow between the scales is analogous to Prigogine's (Prigogine and Stengers, 1988) concept of the spontaneous emergence of order and organization out of apparent disorder and chaos through a process of self-organization.

### 2.4.1 Emergence of order and coherence in biology

The problem of emergence of macroscopic variables out of microscopic dynamics is of crucial relevance in biology (Vitiello, 1992). Biological systems rely on a combination of network and the specific elements involved (Kitano, 2002). The notion that membership in a network could confer stability emerged from Ludwig von Bertalanffy's description of general systems theory in the 1930s and Norbert Wieners description of cybernetics in the 1940s. General systems theory focused in part on the notion of flow, postulating the existence and significance of flow equilibria. In contrast to Cannon's concept that mechanisms should yield homeostasis, general systems theory invited biologists to consider an alternative model of homeodynamics in which nonlinear, non-equilibrium processes could provide stability, if not constancy (Buchman, 2002).

The cell dynamical system model for coherent pattern formation in turbulent flows summarized earlier (Section 2) may provide a general systems theory for biological complexity. General systems theory is a logical-mathematical field, the subject matter of which is the formulation and deduction of those principles which are valid for 'systems' in general, whatever the nature of their component elements or the relations or 'forces' between them (Bertalanffy, 1968; Peacocke, 1989; Klir, 1993).

## 2.5 Long-range spatiotemporal correlations (coherence)

The logarithmic spiral flow pattern enclosing the vortices $OR_OR_1$, $OR_1R_2$, … may be visualized as a continuous smooth rotation of the phase angle $\theta$ ($R_OOR_1$, $R_OOR_2$, … etc.) with increase in period. The phase angle $\theta$ for each stage of growth is equal to $r/R$ and is proportional to the variance $W^2$ (Eq.1), the variance representing the intensity of fluctuations. The phase angle gives a measure of coherence or correlation in space-time fluctuations. The model predicted continuous smooth rotation of phase angle with increase in period length associated with logarithmic spiral flow structure is analogous to Berry's phase (Berry, 1988; Kepler *et al.*, 1991) in quantum systems.

## 2.6 Universal spectrum of fluctuations

Conventional power spectral analysis will resolve such a logarithmic spiral flow trajectory as a continuum of eddies (broadband spectrum) with a progressive increase in phase angle. The power spectrum, plotted on log-log scale as variance versus frequency (period) will represent the probability density corresponding to normalized standard deviation $t$ given by

$$t = \frac{\log L}{\log T_{50}} - 1 \tag{3}$$

In the above Eq. (3) $L$ is the period in years and $T_{50}$ is the period up to which the cumulative percentage contribution to total variance is equal to 50. The above expression for normalized standard deviation $t$ follows from model prediction of logarithmic spiral flow structure and model concept of successive growth structures by space-time averaging.

The period (or length scale) $T_{50}$ up to which the cumulative percentage contribution to total variances is equal to 50 is computed from model concepts as follows

$$T_{50} = (2 + \tau)\tau^0 \tag{4}$$

Fluctuations of all scales therefore self-organize to form the universal inverse power law form of the statistical normal distribution. Since the phase angle θ equal to $r/R$ represents the variance $W^2$ (Eq.1), the phase spectrum plotted similar to variance spectrum will also follow the statistical normal distribution.

## 2.7 Quantum mechanics for subatomic dynamics: apparent paradoxes

The following apparent paradoxes found in the subatomic dynamics of quantum systems (Maddox, 1988) are consistent in the context of atmospheric flows as explained in the following.

### 2.7.1 Wave-particle duality

A quantum system behaves as a wave on some occasions and as a particle at other times. Wave-particle duality is consistent in the context of atmospheric waves, which generate particle-like clouds in a row because of formation of clouds in updrafts and dissipation of clouds in adjacent downdrafts characterizing wave motion (Fig.2).

### 2.7.2 Non-local connection

The separated parts of a quantum system respond as a unified whole to local perturbations. Non-local connection is implicit to atmospheric flow structure quantified in Eq.(1) as ordered two-way energy flow between larger and smaller scales and seen as long-range space-time correlations, namely self-organized criticality. Atmospheric flows self-organize to form a unified network with the quasiperiodic Penrose tiling pattern for internal structure (Fig.1), which provide long-range (non-local) space-time connections.

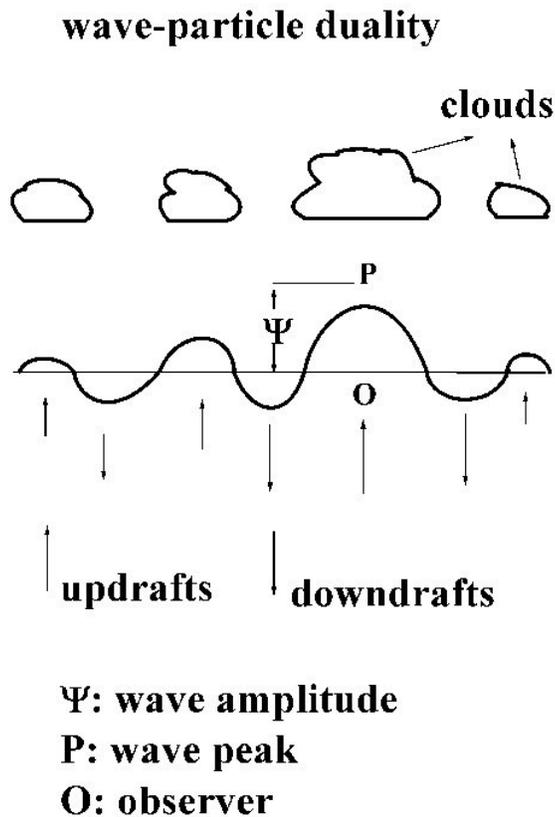

**wave-particle duality**

**clouds**

**P**

**Ψ**

**O**

**updrafts** | **downdrafts**

**Ψ: wave amplitude**
**P: wave peak**
**O: observer**

**wave-trains in atmospheric flows**
**and cloud formation**

Figure 2: Illustration of wave-particle duality as physically consistent for quantumlike mechanics in atmospheric flows. Particlelike clouds form in a row because of condensation of water vapour in updrafts and evaporation of condensed water in adjacent downdrafts associated with eddy circulations in atmospheric flows. Wave-particle duality in macroscale real world dynamical systems may be associated with bimodal (formation and dissipation) phenomenological form for manifestation of energy associated with bidirectional energy flow intrinsic to eddy (wave) circulations in the medium of propagation.

## 3. Applications of the general systems theory concepts to genomic DNA base sequence structure

DNA sequences, the blueprint of all essential genetic information, are polymers consisting of two complementary strands of four types of bases: adenine (A), cytosine (C), guanine (G) and thymine (T). Among the four bases, the presence of A on one strand is always paired with T on the opposite strand, forming a "base pair" with 2 hydrogen bonds. Similarly, G and C are complementary to one another, while forming a base pair with 3 hydrogen bonds. Consequently, one may characterize AT base-pairs as weak bases and GC base-pairs as strong bases. In addition, the frequency of A(G) on a single strand is approximately equal to the frequency of T(C) on the same strand, a phenomenon that has been termed "strand symmetry" or "Chargaff's second parity". Therefore, DNA sequences can be transformed into sequences of weak W (A or T) and strong S (G or C) bases (Li and Holste, 2004). The SW mapping rule is particularly appropriate to analyze genome-wide correlations; this rule corresponds to the most fundamental partitioning of the four bases into their natural pairs in the double helix (G+C, A+T). The composition of base pairs, or GC level, is thus a strand-independent property of a DNA molecule and is related to important physico-chemical properties of the chain (Bernaola-Galvan et al., 2002). The C+G content (isochore) studies have been done earlier (Bernardi, et al., 1985; Ikemura, 1985.; Ikemura and Aota, 1988; Bernardi, G., 1989). The full story of how DNA really functions is not merely what is written on the sequence of base pairs. The DNA functions involve information transmission over many length scales ranging from a few to several hundred nanometers (Ball, 2003).

One of the major goals in DNA sequence analysis is to gain an understanding of the overall organization of the genome, in particular, to analyze the properties of the DNA string itself. Long-range correlations in DNA base sequence structure, which give rise to $1/f$ spectra have been identified (Fukushima et al., 2002; Azad et al., 2002). Such long-range correlations in space-time fluctuations is very common in nature and Li (2007) has given an extensive and informative bibliography of the observed $1/f$ noise or $1/f$ spectra, where $f$ is the frequency, in biological, physical, chemical and other dynamical systems. The long-range correlations in nucleotide sequence could in principle be explained by the coexistence of many different length scales. The advantage of spectral analysis is to reveal patterns hidden in a direct correlation function. The quality of the $1/f$ spectra differs greatly among sequences. Different DNA sequences do not exhibit the same power spectrum.

The concentration of genes is correlated with the C+G density. The spatial distribution of C+G density can be used to give an indication of the location of genes. The final goal is to eventually learn the 'genome organization principles' (Li, 1997). The coding sequences of most vertebrate genes are split into segments (exons) which are separated by noncoding intervening sequences (introns). A very small minority of human genes lack noncoding introns and are very small genes (Strachan and Read, 1996).

Li (2002) reports that spectral analysis shows that there are GC content fluctuations at different length scales in isochore (relatively homogeneous) sequences. Fluctuations of all size scales coexist in a hierarchy of domains within domains (Li et al., 2003). Li and Holste (2005) have recently identified universal $1/f$ spectra and diverse correlation structures in Guanine (G) and Cytosine (C) content of all human chromosomes.

In the following it is shown that the frequency distribution of Human chromosome 1 TO 24 DNA bases C+G concentration per 10bp (non-overlapping) follows the model prediction (Section 2) of self-organized criticality or quantumlike chaos implying long-range spatial correlations in the distribution of bases C+G along the DNA base sequence.

## 4. Data and Analysis

### 4.1 Data

The Human chromosomes 1 to 24 DNA base sequence was obtained from the entrez Databases, Homo sapiens Genome (build 36 Version 1) at http://www.ncbi.nlm.nih.gov/entrez. The total number of contiguous data sets, each containing a minimum of 70 000 base pairs, chosen for the study are given in Fig. 3 for the chromosomes 1 to 24.

### 4.2 Power spectral analyses: variance and phase spectra

The number of times base C and also base G, i.e., (C+G), occur in successive blocks of 10 bases were determined in successive length sections of 70000 base pairs giving a C+G frequency distribution series of 7000 values for each data set. The power spectra of frequency distribution of C+G bases (per 10bp) in the data sets were computed accurately by an elementary, but very powerful method of analysis developed by Jenkinson (1977) which provides a quasi-continuous form of the classical periodogram allowing systematic allocation of the total variance and degrees of freedom of the data series to logarithmically spaced elements of the frequency range (0.5, 0). The cumulative percentage contribution to total variance was computed starting from the high frequency side of the spectrum. The power spectra were plotted as cumulative percentage contribution to total variance versus the normalized standard deviation $t$ equal to $\left(\log L/\log T_{50}\right)-1$ where $L$ is the period in years and $T_{50}$ is the period up to which the cumulative percentage contribution to total variance is equal to *50* (Eq. (3)). The corresponding phase spectra were computed as the cumulative percentage contribution to total rotation (Section 2.6). The statistical chi-square test (Spiegel, 1961) was applied to determine the 'goodness of fit' of variance and phase spectra with statistical normal distribution. Details of data sets and results of power spectral analyses are given in Fig. 3 as averages for each of the 24 chromosomes. The average variance and phase spectra for the data sets in each of the 24 chromosomes (Fig. 3) are given in Fig. 4.

### 4.3 Power spectral analyses: dominant periodicities

The general systems theory predicts the broadband power spectrum of fractal fluctuations will have embedded dominant wavebands, the bandwidth increasing with wavelength, and the wavelengths being functions of the golden mean (Eq.2). The first 13 values of the model predicted (Selvam, 1990; Selvam and Fadnavis, 1998) dominant peak wavelengths are 2.2, 3.6, 5.8, 9.5, 15.3, 24.8, 40.1, 64.9, 105.0, 167.0, 275, 445.0 and 720 in units of the block length 10bp (base pairs) in the present study. The dominant peak wavelengths were grouped into 13 class intervals 2 - 3, 3 - 4, 4 - 6, 6 - 12, 12 - 20, 20 - 30, 30 - 50, 50 - 80, 80 − 120, 120 − 200, 200 − 300, 300 − 600, 600 - 1000 (in units of 10bp block lengths) to include the model predicted dominant peak length scales mentioned above. The class intervals increase in size progressively to accommodate model predicted increase in bandwidth associated with increasing wavelength. Average class interval-wise percentage frequencies of occurrence of dominant wavelengths (normalized variance greater than 1) are shown in Fig. 5 along with the percentage contribution to total variance in each class interval corresponding to the normalised standard deviation $t$ (Eq. 3) computed from the average $T_{50}$ (Fig. 3) for each of the 24 chromosomes. In this context it may be mentioned that statistical normal probability density distribution represents the eddy variance (Eq. 3). The observed frequency distribution of dominant eddies follow closely the computed percentage contribution to total variance.

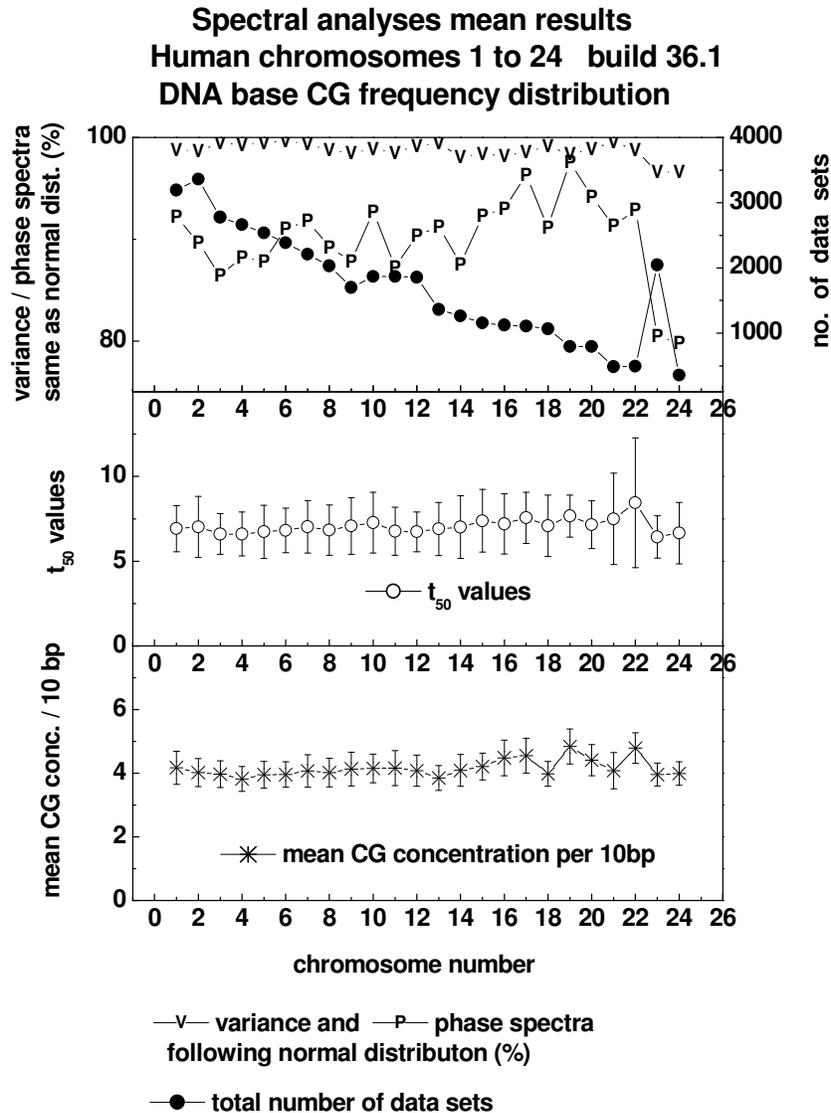

**Spectral analyses mean results**
**Human chromosomes 1 to 24   build 36.1**
**DNA base CG frequency distribution**

Figure 3: Average results of power spectral analyses. The error bars for one standard deviation are given for $t_{50}$ values and mean CG concentration per 10bp

### 4.3.1 Peak wavelength versus bandwidth

The model predicts that the apparently irregular *fractal* fluctuations contribute to the ordered growth of the quasiperiodic *Penrose* tiling pattern with an overall logarithmic spiral trajectory such that the successive radii lengths follow the *Fibonacci* mathematical series. Conventional power spectral analyses resolves such a spiral trajectory as an eddy continuum with embedded dominant wavebands, the bandwidth increasing with wavelength. The progressive increase in the radius of the spiral trajectory generates the eddy bandwidth proportional to the increment $d\theta$ in phase angle equal to $r/R$ The relative eddy circulation speed $W/w_*$ is directly proportional to the relative peak wavelength ratio $R/r$ since the eddy circulation speed $W = 2\pi R/T$ where $T$ is the eddy time period. The relationship between the peak wavelength and the bandwidth is obtained from Eq. (1), namely,

$$W^2 = \frac{2}{\pi}\frac{r}{R}w_*^2 .$$

Considering eddy growth with overall logarithmic spiral trajectory

$$\text{relative eddy bandwidth} \propto d\theta \propto \frac{r}{R}$$

The eddy circulation speed is related to eddy radius as

$$W = \frac{2\pi R}{T}$$

$$W \propto R \propto \text{peak wavelength}$$

The relative peak wavelength is given in terms of eddy circulation speed as

$$\text{relative peak wavelength} \propto \frac{W}{w_*}$$

From Eq. (1) the relationship between eddy bandwidth and peak wavelength is obtained as

$$\text{eddy bandwidth} = (\text{peak wavelength})^2$$

$$\frac{\log(\text{eddy bandwidth})}{\log((\text{peak wavelength})} = 2 \qquad (9)$$

A log-log plot of peak wavelength versus bandwidth will be a straight line with a slope (bandwidth/peak wavelength) equal to 2. A log-log plot of the average values of bandwidth versus peak wavelength shown in Figs. 6 exhibit average slopes approximately equal to 2.5.

**Average variance and phase spectra**
**Human chromosome (1 to 24) DNA base C+G frequency spectra**

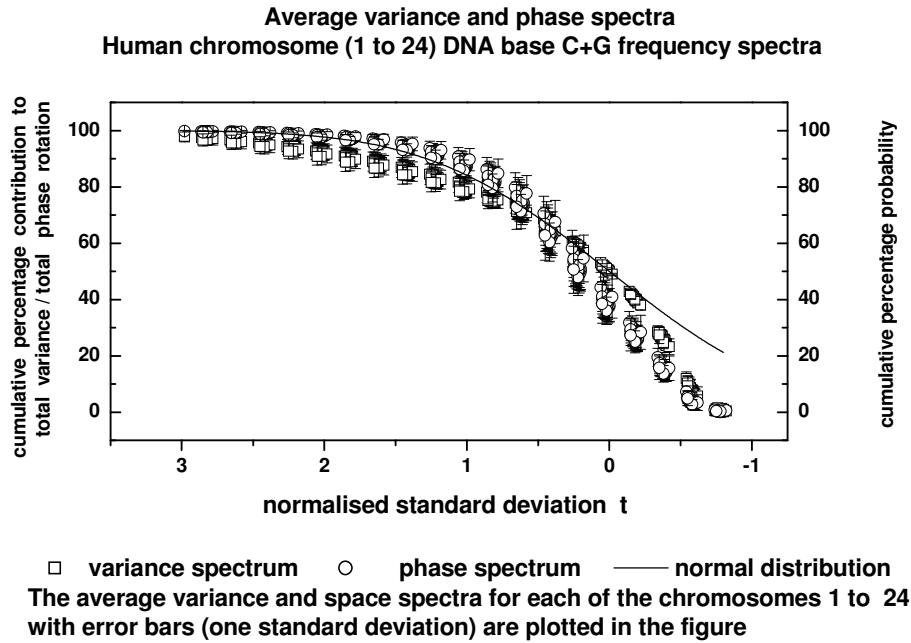

☐  **variance spectrum**      ○  **phase spectrum**      ——  **normal distribution**
**The average variance and space spectra for each of the chromosomes 1 to 24**
**with error bars (one standard deviation) are plotted in the figure**

Figure 4: The average variance and phase spectra of frequency distribution of bases C+G in
Human chromosomes 1 to 24 for the data sets given in Fig. 3. The power spectra
were computed as cumulative percentage contribution to total variance versus the
normalized standard deviation $t$ equal to $(\log L / \log T_{50})$ −1 where $L$ is the period in
years and $T_{50}$ is the period up to which the cumulative percentage contribution to
total variance is equal to 50. The corresponding phase spectra were computed as
the cumulative percentage contribution to total rotation (Section 2.6).

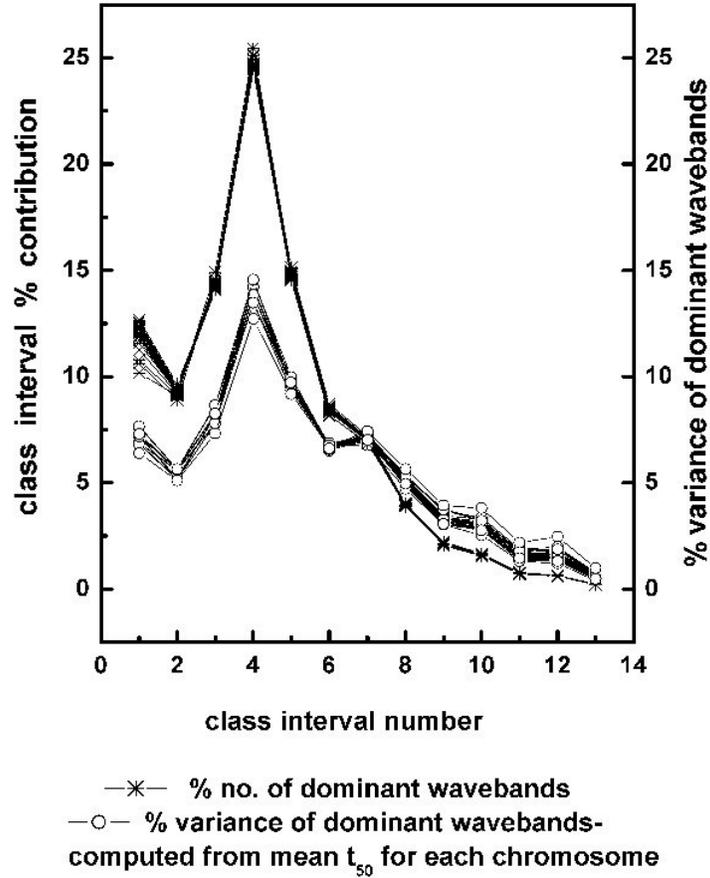

Figure 5: Dominant wavelengths in DNA bases C+G concentration distribution. Average class interval-wise percentage frequency distribution of dominant (normalized variance greater than 1) wavelengths is given by *line + star*. The corresponding computed percentage contribution to the total variance for each class interval is given by *line + open circle*. The observed frequency distribution of dominant eddies closely follow the model predicted computed percentage contribution to total variance.

**Spectral analyses of DNA base C+G sequence
Human chromosomes 1 to 24 build 36.1
Dominant wavebands versus peak wavelength**

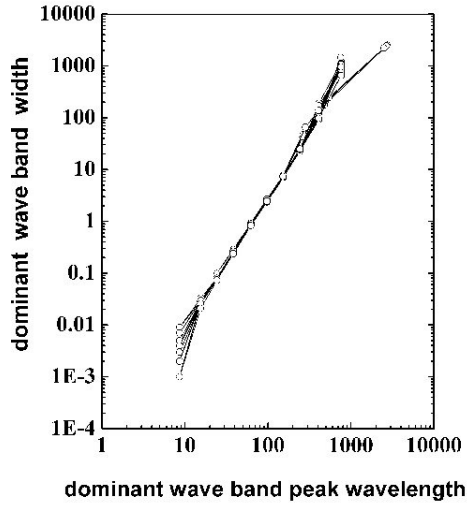

X and Y axes are on log-log scale
—○— bandwidth versus peak wavelength

**Spectral analyses of DNA base C+G sequence
Human chromosomes 1 to 24 build 36.1
Dominant wavebands versus peak wavelength**

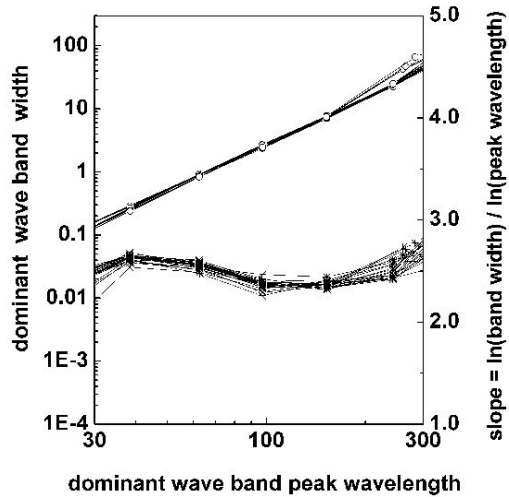

X and Y axes are on log-log scale
—○— bandwidth versus peak wavelength

—✕— slope versus peak wavelength

Figures 6a & 6b: The linear relationship between logarithms of dominant wave bandwidth versus corresponding peak wavelength. Fig 6b (below) shows the slope for a limited range of peak wavelength

## 5. Discussions

In summary, a majority of the data sets (Fig. 3) exhibit the model predicted quantumlike chaos for fractal fluctuations since the variance and phase spectra follow each other closely and also follow the universal inverse power law form of the statistical normal distribution signifying long-range correlations or coherence in the overall frequency distribution pattern of the DNA bases C+G in Human chromosomes 1 to 24. Such non-local connections or 'memory' in the spatial pattern is a natural consequence of the model predicted Fibonacci spiral enclosing the space filling quasicrystalline structure of the quasiperiodic Penrose tiling pattern for fractal fluctuations of dynamical systems. Further, the broadband power spectra exhibit dominant wavelengths closely corresponding to the model predicted (Fig. 1, Eq. 2 and Fig. 4) nested continuum of eddies. The apparently chaotic fluctuations of the frequency distribution of the DNA bases C+G per 10bp in the Human chromosomes 1 to 24 self-organize to form an ordered hierarchy of spirals or loops.

Analysis of auto-correlations of human chromosomes 1–22 and rice chromosomes 1–12 for seven binary mapping rules shows that the correlation patterns are different for different rules but almost identical for all of the chromosomes, despite their varying lengths (Podobnik et al.**,** 2007**)**

Mansilla et al., (2004) calculated the mutual information function for each of the 24 chromosomes in the human genome. The same correlation pattern is observed regardless the individual functional features of each chromosome. Moreover, correlations of different scale length are detected depicting a multifractal scenario. This fact suggest a unique mechanism of structural evolution.

Quasicrystalline structure of the quasiperiodic Penrose tiling pattern has maximum packing efficiency as displayed in plant phyllotaxis (Selvam, 1998) and may be the geometrical structure underlying the packing of $10^3$ to $10^5$ micrometer of DNA in a eukaryotic (higher organism) chromosome into a metaphase structure (before cell division) a few microns long as explained in the following. A length of DNA equal to $2\pi L$ when coiled in a loop of radius $L$ has a packing efficiency (lengthwise) equal to $2\pi L/2L=\pi$ since the linear length $2\pi L$ is now accommodated in a length equal to the diameter $2L$ of the loop. Since each stage of looping gives a packing efficiency equal to $\pi$, ten stages of such successive looping will result in a packing efficiency equal to $\pi^{10}$ approximately equal to $10^5$.

The present study deals with all the 24 human chromosome bases C+G concentration per 10bp in all available contiguous sequences. The window length 10bp was chosen since the primary loop in the DNA molecule is equal to about 10bp. The power spectral analysis gives the dominant wavelengths in terms of this basic unit, namely the window length of 10bp. Increasing the window length (more than 10bp) will result in decrease in resolution of shorter wavelengths. The aim of this preliminary study is to determine the spatial organization of the DNA bases C+G by applying concepts of a general systems theory first developed for atmospheric flows.

The important results of the present study are as follows: (1) the concentration per 10bp of DNA bases C+G follow selfsimilar fractal fluctuations, namely an irregular series of successive increase followed by decrease on all size scales. (2) The power spectra of C+G concentration distribution follow the inverse power law form of the statistical normal distribution signifying quasicrystalline structure of the quasiperiodic Penrose tiling pattern for the spatial distribution of DNA bases C+G. (3) The quasiperiodic Penrose tiling pattern provides maximum packing efficiency for the DNA molecule inside the chromosome. (4) The observed inverse power law form for power spectra implies that the DNA bases are arranged in a fuzzy logic network with inherent long-range correlations.

## 6. Conclusions

Real world and model dynamical systems exhibit long-range space-time correlations, i.e., coherence, recently identified as self-organized criticality. Macroscale coherent functions in biological systems develop from self-organization of microscopic scale information flow and control such as in the neural networks of the human brain and in the His-Purkinje fibers of human heart, which govern vital physiological functions.

A recently developed cell dynamical system model for turbulent fluid flows predicts self-organized criticality as intrinsic to quantumlike mechanics governing flow dynamics. The model concepts are independent of exact details (physical, chemical, biological etc.) of the dynamical system and are universally applicable. The model is based on the simple concept that space-time integration of microscopic domain fluctuations occur on selfsimilar fractal structures and give rise to the observed space-time coherent behaviour pattern with implicit long-term memory. Selfsimilar fractal structures to the spatial pattern for dynamical systems function as dynamic memory storage device with memory recall and update at all time scales.

The important conclusions of this study are as follows: (1) the frequency distribution of bases C+G per 10bp in all the 24 human chromosomes DNA exhibit selfsimilar fractal fluctuations which follow the universal inverse power law form of the statistical normal distribution (Fig. 4), a signature of quantumlike chaos. (2) Quantumlike chaos indicates long-range spatial correlations or 'memory' inherent to the self-organized fuzzy logic network of the quasiperiodic Penrose tiling pattern (Eq. 1 and Fig. 1). (3) Such non-local connections indicate that coding exons together with non-coding introns contribute to the effective functioning of the DNA molecule as a unified whole. Recent studies indicate that non-coding DNA introduce modifications in gene activity (Cohen, 2002; Makalowski, 2003). Studies now indicate that non-coding DNA may be responsible for the signals that were crucial for human evolution, directing the various components of our genome to work differently from the way they do in other organisms (Check, 2006). (4) The space filling quasiperiodic Penrose tiling pattern provides maximum packing efficiency for the DNA molecule inside the chromosome.

## Acknowledgement

The author is grateful to Dr. A. S. R. Murty for encouragement.